\documentclass[
  journal=pasa,
  manuscript=article-type,
  year=2025,
  volume=37,
]{cup-journal}

\usepackage{amsmath}
\usepackage[nopatch]{microtype}
\usepackage{booktabs}
\usepackage{amsmath,amssymb}
\usepackage{graphics}
\usepackage{hyperref}
\usepackage{multirow}

\hypersetup{colorlinks=true,linkcolor=blue,filecolor=blue,urlcolor=blue,citecolor=blue}

\usepackage{natbib}
\newcommand{\kms}{km~s$^{-1}$}

\newcommand{\msunpc}{$\rm M_\odot~pc^{-2}$}
\newcommand{{\hi}}{{H{\sc i}}}

\newcommand{\rband}{{\em r}-band}
\newcommand{\gband}{{\em g}-band}
\newcommand{\iband}{{\em i}-band}

\newcommand{\Bband}{{\em B}-band}

\newcommand{\Mhi}{$M_{\rm HI}$}
\newcommand{\Mst}{$M_\star$}

\newcommand{\gi}{$g_{\rm 25}-i_{\rm 25}$}

\newcommand{\riso}{$R_{\rm 25}$}
\newcommand{\mycomment}[1]{}

\title{WALLABY pilot survey: Spatially resolved gas scaling relations within the stellar discs of nearby galaxies}

\author{Seona Lee}
\affiliation{International Centre for Radio Astronomy Research (ICRAR), The University of Western Australia, Crawley, WA, Australia}
\alsoaffiliation{ARC Centre of Excellence for All Sky
Astrophysics in 3 Dimensions (ASTRO 3D), Australia}
\email[Seona Lee]{seona.lee@icrar.org}

\author{Barbara Catinella}
\affiliation{International Centre for Radio Astronomy Research (ICRAR), The University of Western Australia, Crawley, WA, Australia}
\alsoaffiliation{ARC Centre of Excellence for All Sky
Astrophysics in 3 Dimensions (ASTRO 3D), Australia}
\author{Tobias Westmeier}
\affiliation{International Centre for Radio Astronomy Research (ICRAR), The University of Western Australia, Crawley, WA, Australia}
\alsoaffiliation{ARC Centre of Excellence for All Sky
Astrophysics in 3 Dimensions (ASTRO 3D), Australia}

\author{Luca Cortese}
\affiliation{International Centre for Radio Astronomy Research (ICRAR), The University of Western Australia, Crawley, WA, Australia}
\alsoaffiliation{ARC Centre of Excellence for All Sky
Astrophysics in 3 Dimensions (ASTRO 3D), Australia}

\author{Jing Wang}
\affiliation{Kavli Institute for Astronomy and Astrophysics, Peking University, Beijing, China}

\author{Kristine Spekkens}
\affiliation{Department of Physics, Engineering Physics, and Astronomy, Queen’s University, Kingston, ON, Canada}

\author{Nathan Deg}
\affiliation{Department of Physics, Engineering Physics, and Astronomy, Queen’s University, Kingston, ON, Canada}

\author{Helga Dénes}
\affiliation{School of Physical Sciences and Nanotechnology, Yachay Tech University, Urcuquí, Ecuador}

\author{Ahmed Elagali}
\affiliation{School of Biological Sciences, The University of
Western Australia, Perth, WA, Australia}

\author{B\"arbel S. Koribalski}
\affiliation{Australia Telescope National Facility, CSIRO, Space and Astronomy, Epping, NSW, Australia}
\alsoaffiliation{School of Science, Western Sydney University, Penrith, NSW, Australia}

\author{Karen Lee-Waddell}
\affiliation{Australian SKA Regional Centre (AusSRC) – The University of Western Australia, Crawley, WA, Australia}
\alsoaffiliation{Australia Telescope National Facility, CSIRO, Space and Astronomy, Bentley, WA, Australia}
\alsoaffiliation{International Centre for Radio Astronomy Research (ICRAR), Curtin University, Bentley, WA, Australia }

\author{Chandrashekar Murugeshan}
\affiliation{Australia Telescope National Facility, CSIRO, Space and Astronomy, Epping, NSW, Australia}
\alsoaffiliation{ARC Centre of Excellence for All Sky Astrophysics in 3 Dimensions (ASTRO 3D), Australia}

\author{Jonghwan Rhee}
\affiliation{International Centre for Radio Astronomy Research (ICRAR), The University of Western Australia, Crawley, WA, Australia}

\author{Lister Staveley-Smith}
\affiliation{International Centre for Radio Astronomy Research (ICRAR), The University of Western Australia, Crawley, WA, Australia}
\alsoaffiliation{ARC Centre of Excellence for All Sky Astrophysics in 3 Dimensions (ASTRO 3D), Australia}

\author{O. Ivy Wong}
\affiliation{Australia Telescope National Facility, CSIRO, Space and Astronomy, Bentley, WA, Australia}
\alsoaffiliation{International Centre for Radio Astronomy Research (ICRAR), The University of Western Australia, Crawley, WA, Australia}
\alsoaffiliation{ARC Centre of Excellence for All Sky Astrophysics in 3 Dimensions (ASTRO 3D), Australia}

\author{Benne W. Holwerda}
\affiliation{Department of Physics and Astronomy, University of Louisville, Louisville, KY, USA}

 %% First letter not capped

\begin{document}

\begin{abstract}

The scatter in global atomic hydrogen (\hi) scaling relations is partly attributed to differences in how \hi\ and stellar properties are measured, with \hi\ reservoirs typically extending beyond the inner regions of galaxies where star formation occurs. Using pilot observations from the Widefield ASKAP L-band Legacy All-sky Blind Survey (WALLABY), we present the first measurements of \hi\ mass enclosed within the stellar-dominated regions of galaxies for a statistical sample of 995 local gas-rich systems, investigating the factors driving its variation.

We examine how global \hi\ scaling relations change when measurements are restricted to $R_{\rm 25}$ and $R_{\rm 24}$ -- the isophotal radii at 25 and 24 mag arcsec$^{-2}$ in the \iband\ -- and explore how the fraction of \hi\ mass and \hi\ surface density within these radii correlate with other galaxy properties. On average, 68\% of the total \hi\ mass is enclosed within $R_{\rm 25}$ and 54\% within $R_{\rm 24}$, though significant variation exists between galaxies, ranging from $\sim$20\% to 100\%. The fraction of \hi\ mass within $R_{\rm 25}$ shows a mild correlation with stellar properties, with galaxies of higher stellar mass, greater stellar surface density, or redder colours enclosing a larger fraction of their \hi\ reservoirs. These correlations do not significantly strengthen when considering $R_{\rm 24}$.
Conversely, global \hi\ surface densities show no significant correlation with stellar mass or stellar surface density, but trends start emerging when these are measured within the inner regions of galaxies. The strongest correlation is observed with optical colour, with bluer galaxies having higher average \hi\ surface densities within $R_{\rm 25}$. This trend of the average \hi\ surface density with optical colour strengthens when we restrict from $R_{\rm 25}$ to $R_{\rm 24}$, suggesting a closer connection between inner \hi\ reservoirs and star formation. This study underscores the value of (at least marginally) resolved \hi\ surveys of statistical samples for advancing our understanding of the gas-star formation cycle in galaxies.

\end{abstract}

\section{Introduction}

Neutral atomic hydrogen ({\hi}) plays a crucial role in galaxy evolution by serving as the primary source of cold gas that fuels star formation. 
Traditionally, {\hi} observations using single-dish radio telescopes have provided global {\hi} data for large galaxy samples \citep[e.g. the Arecibo Legacy Fast ALFA survey, ALFALFA;][]{Giovanelli2005,Haynes2018}, offering insights into how galaxies consume their gas content to sustain star formation \citep[e.g.,][and references therein]{Saintonge2022}.
For example, there is an inverse relation between a galaxy's gas-to-stellar mass fraction and stellar quantities, such that early-type galaxies, typically more massive and bulge-dominated, tend to have lower gas mass fractions.
The gas mass fraction correlates most strongly with star formation quantities, which indicates that gas-rich galaxies tend to be actively forming stars \citep[e.g.,][]{Saintonge2022}.
However, {\hi} is converted into molecular gas in dense regions with surface densities exceeding about 10 {\msunpc} \citep{Martin2001,Bigiel2008,Bigiel2012}, becoming a more direct precursor to form stars in galaxies.
Indeed, {\hi} extends well beyond a galaxy's stellar disc—typically twice as far—and star formation in these outer {\hi} regions is less efficient than in the central galactic disc \citep{Bigiel2010a}.
The limited spatial resolution of single-dish telescopes has hindered our ability to resolve detailed {\hi} structures within galaxies, thus limiting our understanding of its role on similar scales with other galactic properties.

Recent advancements in radio interferometry have provided the ability to study {\hi} properties with improved spatial resolution.
For example, several studies have identified a close relation between {\hi} isodensity radius and {\hi} mass \citep[e.g.][]{Broeils1997,Swaters2002,Wang2016,Rajohnson2022}.
In particular, \citet{Wang2016} combined 15 different {\hi} interferometric surveys and demonstrated a remarkably low scatter of this relation.
Using analytical {\hi} models, \citet{Stevens2019} further linked these findings to the universal distribution of {\hi} in the outer regions of galaxies, typically described by a declining exponential function \citep{Swaters2002,Wang2014}.
However, studies have observed deviations from the universal distribution in inner {\hi} radial profiles, often showing flattened or damped {\hi} surface densities near the galactic centre \citep[e.g.][]{Swaters2002,Leroy2008,Bigiel2010a,Wang2014}.
They found that the central surface density of {\hi} varies widely, ranging from 1 to $\sim$10 {\msunpc}. However, how other galaxy properties influence this variation remains unclear, mainly because of the small sample sizes.
Despite this progress, understanding the variation of {\hi} content within galaxies on smaller scales has been challenging due to the limited availability of homogeneous large samples with spatially resolved {\hi} data.

Previous studies investigated spatially resolved \hi\ scaling relations, finding weak or no correlation between {\hi} surface density and star formation surface density on $\sim$kpc scales for small samples of very nearby galaxies \citep[e.g.][]{Wong2002,Boissier2003,Bigiel2008,Schruba2011,Watts2023}.
When focusing on \hi\ within the optical radius, \citet{Wang2017} reported a very weak correlation between average \hi\ and SFR surface densities for galaxies from the Local Volume \hi\ Survey \citep[LVHIS;][]{Koribalski2018}, while \citet{Naluminsa2021} found no correlation for galaxies from the Westerbork survey of \hi\ in Irregular and Spiral galaxies \citep[WHISP;][]{van_der_Hulst2001}.
To conduct a more extensive analysis, \citet{Wang2020} introduced a method to estimate the {\hi} mass within the optical radius from global {\hi} spectra, using the {\hi} mass-size relation and median {\hi} radial profiles derived from 168 spatially resolved {\hi} maps.
Applying their technique to 447 late-type galaxies from the extended GALEX Arecibo SDSS Survey \citep[xGASS;][]{Catinella2018}, they showed similar correlations in {\hi} mass fraction scaling relations but with a shallower slope and decreased scatter.
However, their findings were based on estimates derived from unresolved data, which require further validation through observations.

The Widefield ASKAP L-band Legacy All-sky Blind Survey \citep[WALLABY;][]{Koribalski2020} can address this issue for the first time for a large sample of spatially resolved {\hi} detections in the local Universe. 
Conducted with the Australian Square Kilometre Array Pathfinder \citep[ASKAP;][]{Johnston2008,Hotan2021}, WALLABY is expected to detect over 200,000 {\hi} sources across a significant part of the southern hemisphere out to redshifts of 0.08.
The WALLABY pilot survey has already provided {\hi} source catalogues, images, and spectra of more than 2,000 {\hi} detections \citep{Westmeier2022,Murugeshan2024}.

In this study, we use WALLABY to quantify variations in {\hi} content (mass and surface density) within the stellar disc of 995 galaxies and investigate the causes for these variations.
This paper is structured as follows.
In Section \ref{sec_data} and \ref{sec_method}, we introduce the {\hi} and optical data and outline the measurement of physical quantities. Section \ref{sec_sample} describes our sample selection from WALLABY, and Section \ref{sec_result} presents our findings on {\hi} within stellar disc properties and their relation to stellar properties. Finally, in Section \ref{sec_discussion}, we discuss our results in context with previous studies and conclude in Section \ref{sec_conclusion}.
This paper uses the AB magnitude system and assumes a flat $\Lambda$CDM model with $H_{0}=$ 70 {\kms} Mpc$^{-1}$ \citep{Planck2020}.

\section{Data}
\label{sec_data}
WALLABY provides a statistical sample of spatially-resolved {\hi} data.
The first pilot survey released {\hi} data observed toward the Hydra cluster, the NGC 4636 group and the Norma cluster (60 deg$^2$ each) \citep[PDR1;][]{Westmeier2022}.
WALLABY then released the second pilot survey data observed toward the NGC 5044 group (120 deg$^2$), the NGC 4808 group (30 deg$^2$), and the Vela cluster (30 deg$^2$) \citep[PDR2;][]{Murugeshan2024}.
Among them, we use only data from the Hydra cluster, NGC 4636, NGC 5044 and NGC 4808 fields (272, 147, 1326, and 231 {\hi} detections, respectively) since Norma and Vela fields are affected by strong continuum residuals.
WALLABY provides source catalogues of {\hi} sources detected using the Source Finding Application 2 \citep[SOFIA2;][]{Westmeier2021} and data products of each {\hi} source such as the {\hi} spectral line cube and {\hi} intensity map (moment 0), with a spatial and spectral resolution of 30 arcsec and 4 {\kms}, respectively, at a sensitivity of 1.6 mJy per beam per 4 {\kms} channel.
The mean rms noise levels are 2.0, 2.7, 1.8, and 1.9 mJy per beam for {\hi} detections in the Hydra cluster, NGC 4636, NGC 5044 and NGC 4808 fields, respectively.
Detailed descriptions of WALLABY observations, data processing, and source finding can be found in \citet{Westmeier2022}.
Since WALLABY is an untargeted {\hi} survey, most detections are from gas-rich star-forming galaxies in the local Universe, i.e., z $\lesssim$ 0.08 \citep[e.g., see Figure 1 in][]{Reynolds2023}.

As ancillary data, we use the Dark Energy Spectroscopic Instrument (DESI) Legacy Survey Data Release 10 \citep{Dey2019} to measure the stellar properties of WALLABY sources (e.g. optical radius, stellar mass, stellar surface density, and optical colour).
It includes the Dark Energy Camera Legacy Survey (DECaLS) which provides the sky-subtracted Dark Energy Camera images in multiple optical bands covering both the northern and southern hemispheres (declination < 34$^{\circ}$) with a native pixel scale of 0.262 arcsec.
We obtain {\gband} and {\iband} images using the DESI Legacy Imaging Surveys cutout service by adopting the centroid position of each {\hi} detection from the WALLABY catalogue and the size of the {\hi} intensity map as the input parameters.

We exclude WALLABY \hi\ detections contaminated by nearby radio continuum sources and partial \hi\ detections, resulting in a sample of 1,656 galaxies out of 1,976 initial detections.
We then visually inspect the {\gband} and {\iband} images of the remaining galaxies, selecting only those with no significant contamination from foreground sources or severe background artifacts, and a single optical counterpart.
This narrowed our sample to 1,543 galaxies, 94\% of which have a {\hi} signal-to-noise ratio greater than 5, with a minimum of 3.

\section{Methodology} 
\label{sec_method}

We derive stellar, global {\hi}, and {\hi} within stellar disc properties (e.g., sizes and masses) by conducting photometry on DECalS {\gband} and {\iband} images as well as {\hi} intensity maps.
Instead of selecting an arbitrary stellar radius, we use two isophotal radii at surface brightness levels of 25 and 24 mag arcsec$^{-2}$ in the {\iband} ($R_{\rm 25}$ and $R_{\rm 24}$, respectively).
We measure the {\hi} mass within these two radii, calculate corresponding properties, and analyse whether and how trends or scatter change as we move from global measurements to those at smaller radii.
%\textcolor{red}{ [Explain that, instead of choosing an arbitrary stellar radius, we compute quantities at 2 isophotal radii and investigate whether trends or scatter change when we move from global to smaller radii. See my text in the Discussion.]}

\subsection{Stellar properties}

\begin{figure*}[h]
\centering
\includegraphics[width=\linewidth]{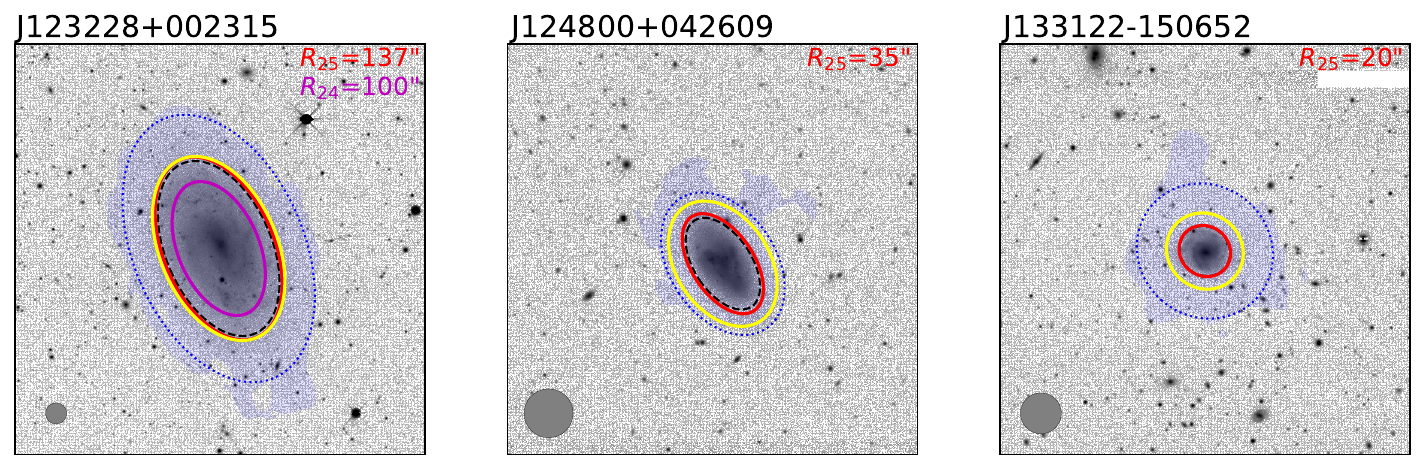}
\caption{DECaLS {\iband} images overlaid with the stellar discs defined by $R_{\rm 25}$ (red solid ellipse), $R_{\rm 24}$ (magenta solid ellipse; shown only in the first panel), and $R_{\rm 90\%}$ (black dashed ellipse) for galaxies with three different resolutions. The yellow solid ellipse represents the convolved stellar disc radius, $R_{\rm 25,c}$ (see text for details).  The blue shaded region shows the \hi\ distribution, with the outer contour corresponding to an integrated {\hi} intensity of 0.4 {\msunpc}, whereas the blue dotted circle indicates the {\hi} disc defined by $R_{\rm HI}$. The value of $R_{\rm 25}$ (and $R_{\rm 24}$ in the first panel) is noted in the top-right corner of each panel. The filled grey circle in the bottom-left corner indicates the 30" WALLABY synthesised beam.}
\label{fig_cartoon_R25}
\end{figure*}

We conduct optical photometry using DECalS {\iband} as the primary reference and extend it to the {\gband}, as the {\iband} is less impacted by dust extinction and better reflects the stellar mass of galaxies.
We measure $R_{\rm 25}$ and $R_{\rm 24}$ isophotal radii.
Previous studies have shown that isophotal radii provide tighter scaling relations compared to radii enclosing a specific fraction of the total light, such as $R_{\rm 50\%}$ (the effective radius) and $R_{\rm 90\%}$ \citep[e.g.][]{Saintonge2011a,Hall2012}.
The scatter in the stellar mass-size relation is known to be minimised when the isophotal radius is defined at a surface brightness of 24.7 mag arcsec$^{-2}$ in the {\rband} \citep[Figure 8 in][]{Sanchez2020}.
Our $R_{\rm 25}$ is effectively equal to $R_{\rm 90\%}$ for the \iband\ in DECaLS imaging, with a mean $R_{\rm 90\%}$/$R_{\rm 25}$ ratio of 0.97 and a standard deviation of 0.26 (see Figure \ref{fig_cartoon_R25}).
This corresponds to a stellar surface density of 1.9 {\msunpc} assuming a mass-to-light ratio of 0.7, which is the average value for our parent sample (see Section \ref{sec_sample}).
In contrast, $R_{\rm 24}$ is closer to the radius at 25 mag arcsec$^{-2}$ in the \Bband\ historically used to define the extent of the optical disc \citep[e.g.,][]{Cortese2012}.

We extract surface brightness profiles to measure $R_{\rm 25}$ and $R_{\rm 24}$ following the method described in \citet{Reynolds2023} but using DECaLS {\iband} images.
In summary, we create a segmentation map to obtain parameters for ellipse fitting and mask sources other than the target galaxy.
We make elliptical annuli and measure the local background level as the sigma-clipped mean image pixel units (ADU) between two ellipses outside the galaxy, i.e., the outermost ellipse and the ellipse with a major axis equal to 75\% of that of the outermost ellipse, from the masked image.
The mean local 3 sigma noise level of DECaLS {\iband} images corresponds to 25.3 mag arcsec$^{-2}$.
After local background subtraction, we measure the mean ADU within each annulus and convert it to surface brightness using
\begin{equation}
    \frac{m}{\mathrm{mag\, arcsec^{-2}}}=22.5-2.5{\rm \,log\frac{ADU}{P_{DECaLS}^2}},
\label{eq_ADU}
\end{equation}
where the DECaLS pixel scale ($\rm P_{DECaLS}$) is 0.262 arcsec per pixel.
We define the stellar disc sizes as the semi-major axis of the isophotal ellipse where the {\iband} surface brightness is 25 or 24 mag arcsec$^{-2}$.

Using the masked and local background subtracted {\iband} and {\gband} images, we estimate the {\iband} ($M_{\rm i,25}$ and $M_{\rm i,24}$) and {\gband} magnitudes ($M_{\rm g,25}$ and $M_{\rm g,24}$) as the total magnitudes enclosed by the $R_{\rm 25}$ and $R_{\rm 24}$ elliptical apertures, respectively, and the asymptotic magnitude in the {\iband} ($M_{\rm i,asymp}$) using the curve-of-growth \citep[e.g.][]{Munoz-Mateos2015}.
The effective {\iband} radius ($R_{\rm 50\%}$) is estimated as the semi-major axis of the aperture enclosing half the flux corresponding to $M_{\rm i,asymp}$.
The {\gband} and {\iband} magnitudes are corrected for galactic extinction from the Milky Way following the dust extinction law in \citet{Cardelli1989} with the extinction coefficients of $R_{v}=3.214$ and $1.592$ for the DECam {\em g} and {\em i} filters, respectively \citep{Dey2019}.

We estimate stellar masses using the relation given in \citet{Taylor2011} as
\begin{equation}
\begin{split}
    {\rm log}{\frac{M_{\rm \star}}{M_{\rm \odot}}}=&-0.68+0.70\left(g_{\rm 25}-i_{\rm 25}\right)+0.4M_{\rm sol}\\
    &-0.4\left(m-5{\rm log}\frac{D_{\rm L}}{\rm Mpc}-25\right),
\label{eq_ML}
\end{split}
\end{equation}
where {\gi} is the extinction-corrected colour measured within the $R_{\rm 25}$ elliptical aperture, $M_{\rm sol}$ = 4.52 is the absolute magnitude of the Sun in the {\iband} \citep{Willmer2018}, $D_{\rm L}$ is the luminosity distance which is taken as the local Hubble distance from the WALLABY source catalogue\footnote{Our sample may contain cluster galaxies \citep[e.g., 34\% of the \hi\ detections in the Hydra field are classified as cluster galaxies in ][]{Reynolds2023}. However, our key results are distance-independent (e.g., {\hi} mass fraction, {\hi} surface density, etc). Hence, more accurate distances would not affect the trends presented in this paper.} \citep{Westmeier2022}, and $m$ is the apparent {\iband} magnitude.
This relation is known to estimate the stellar mass-to-light ratio with a 1$\sigma$ accuracy of $\sim$ 0.01 dex using only {\em g-} and \iband\ photometry \citep{Taylor2011}.
The total stellar mass ({\Mst}) and stellar mass within $R_{\rm 25}$ ($M_{\rm \star,R25}$) are derived using $m$ from $M_{\rm i,asymp}$ and $M_{\rm i,25}$, respectively. 
Stellar masses measured within $R_{\rm 24}$ are very similar (average $M_{\rm \star,R24}/M_{\rm \star,R25}=0.91$ for our primary sample; see Section \ref{sec_sample}), thus for simplicity, we show our results only for $M_{\rm \star,R25}$.
We derive the average stellar mass surface density as
\begin{equation}
    \mu_{\rm \star}=\frac{M_{\rm \star}}{2\pi R_{50\%}^2}.
\end{equation}

\subsection{Global {\hi} and {\hi} within stellar disc properties}

We study the \hi\ surface brightness distribution of the sample using WALLABY moment 0 (intensity) maps, which have been flux-corrected as described in \citet{Westmeier2022} and \citet{Murugeshan2024}.
We extract radial {\hi} surface density profiles from the intensity maps to measure the {\hi} isodensity radius at 1 {\msunpc} ($R_{\rm HI}$) following the method described in \citet{Reynolds2023}.
In summary, we make elliptical annuli, using the parameters from the 2-dimensional Gaussian fitting to the {\hi} intensity map, binned by one-third of the WALLABY beam's full-width half maximum ($\sim$30"/3) along the major axis, and interpolate the average {\hi} surface density within each annulus to produce the radial profile.
We measure $R_{\rm HI}$ as the radius where the {\hi} surface density is 1 {\msunpc} from the profile.
We do not apply inclination correction to the profile.
The highly-inclined galaxies (i.e., $i$ > 80 degrees) correspond to 3.9\% of our parent sample, and applying inclination correction for $R_{\rm HI}$ does not change the interpretation of our results.
Only when we analyse the shape of radial {\hi} surface density profiles in Section \ref{sec_discussion}, we exclude highly-inclined galaxies and use the inclination-corrected {\hi} profiles and deprojected {\hi} radius ($R_{\rm HI,dep}$) where the {\hi} surface density is 1 {\msunpc} from the inclination-corrected profile.

\begin{figure*}[hbt!]
\centering
\includegraphics[width=0.9\linewidth]{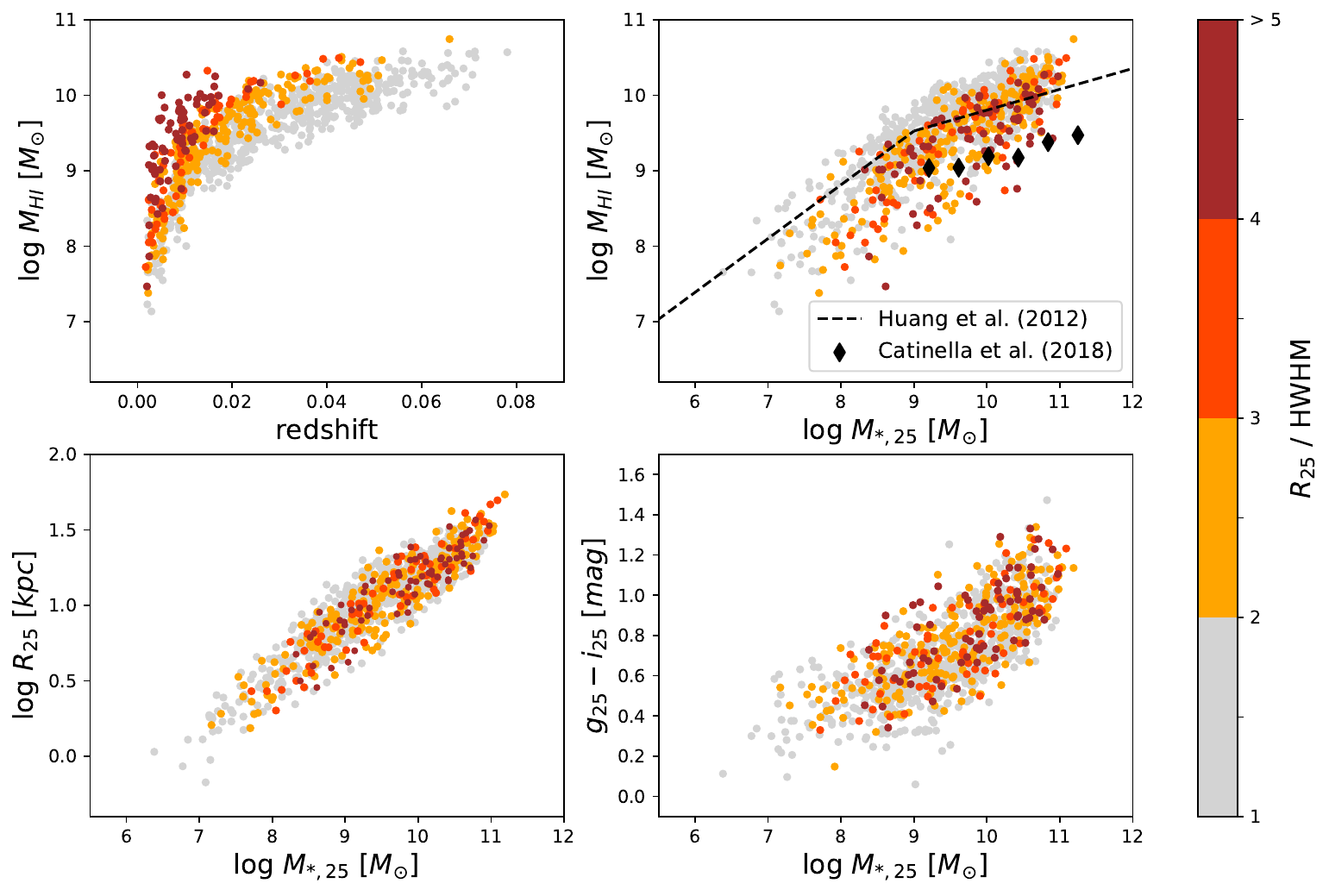}
\caption{Physical properties of our parent sample. Total {\hi} mass as a function of redshift (upper left) and the relations between stellar mass and total \hi\ mass (upper right), stellar radius (lower left), and colour (lower right). The black dashed line is the {\hi}-stellar mass relation derived by \citet{Huang2012} and the black diamonds are the medians of the xGASS galaxies \citep{Catinella2018}. Galaxies are colour-coded from grey to darker colours by the number of beams along the major axis of the stellar disc ($R_{\rm 25}$/HWHM), where HWHM is the WALLABY beam's half-width half maximum (=15").}
\label{fig_sample_RHI}
\end{figure*}

We calculate the total {\hi} mass ($M_{\rm HI}$) by integrating the total {\hi} flux from the {\hi} intensity map and converting it to {\hi} mass \citep[Equation 48 in][]{Meyer2017}.
To estimate the {\hi} mass within $R_{\rm HI}$ ($M_{\rm HI,RHI}$) and within the stellar disc ($M_{\rm HI,R25(24)}$), we use the {\hi} mass curve-of-growth profile, which represents the radial profile of the enclosed {\hi} mass within elliptical apertures.
These apertures are defined based on the central position and position angle derived from the DECaLS {\iband} image, and the axis ratio is obtained from the {\hi} intensity map.
For 86\% of the sample, the separation between the {\hi} and optical centre position is less than 10". We adopt the optical centre for our analysis as it is more likely to provide an accurate reference.
$M_{\rm HI,RHI}$ and $M_{\rm HI,R25(24)}$ are determined using the {\hi} mass curve-of-growth profile at $R_{\rm HI}$ and $R_{\rm 25(24),c}$, where the subscript "c" indicates convolution with the 30" WALLABY synthesised beam to ensure consistent resolution between WALLABY and DECaLS.
In detail, we generate a two-dimensional stellar image based on the {\iband} stellar surface brightness profile, then convolve this image with a Gaussian function representing the WALLABY synthesised beam.
From the convolved image, we re-extract the stellar surface brightness profile and measure $R_{\rm 25(24),c}$.
From this point onward, even if not explicitly stated, the \hi\ mass within stellar discs ($M_{\rm HI,R25(24)}$) and all related properties are assumed to be based on the optical radii convolved to WALLABY resolution.

\begin{figure*}[ht!]
\centering
\includegraphics[width=\linewidth]{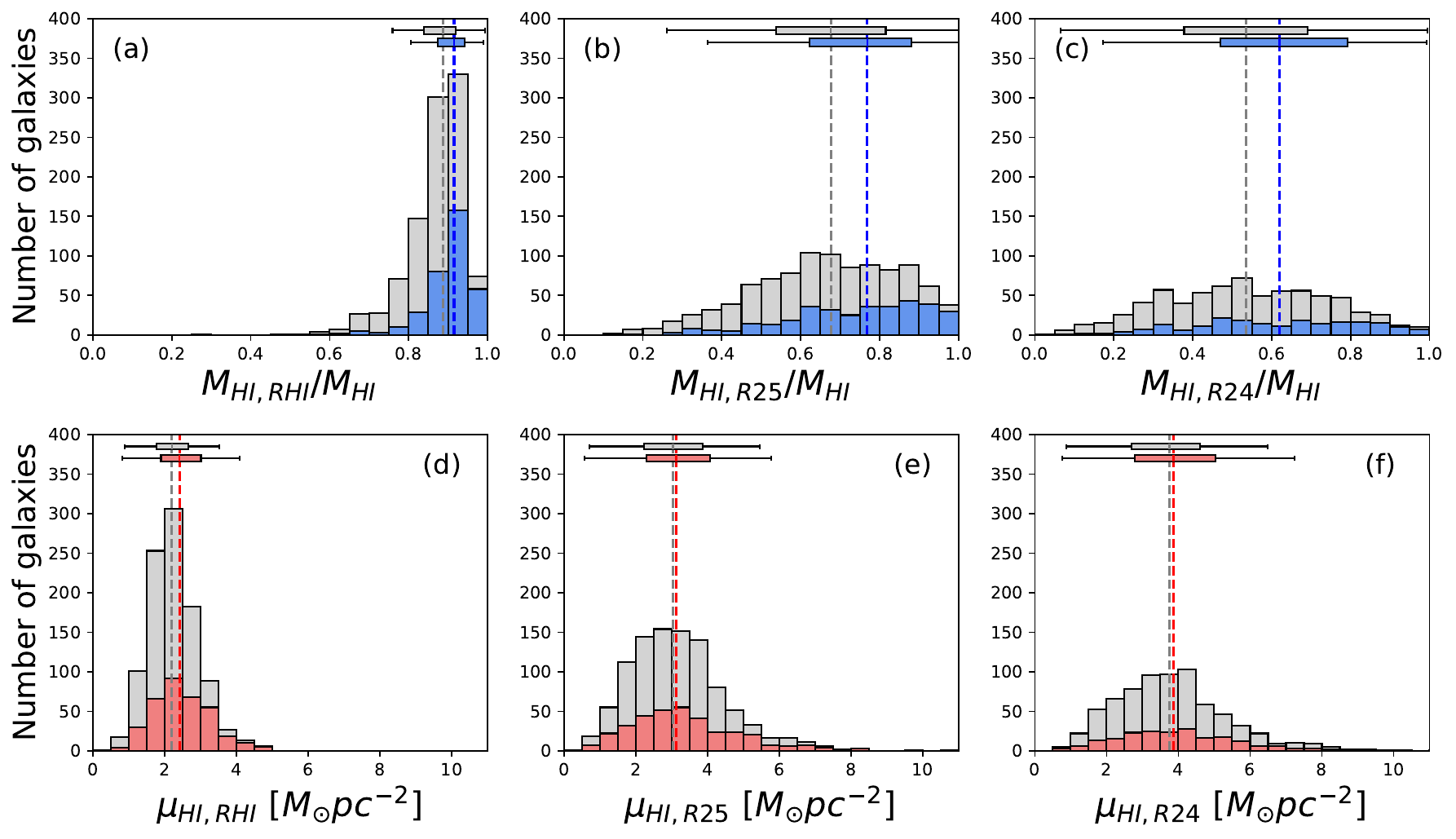}
\caption{Histograms of {\hi} mass enclosed within $R_{\rm HI}$ (a), $R_{\rm 25}$ (b), and $R_{\rm 24}$ (c) normalised by the total {\hi} mass. The bottom panels show the histograms of the average \hi\ surface densities within the same radii (d, e, and f). In all panels, grey and coloured distributions refer to full and higher-resolution samples, dashed lines indicate medians and whisker box plots show the median $\pm$ the interquartile range. Quantities involving $R_{\rm 24}$ are computed using the primary sample.}
\label{fig_hist_HI_in}
\end{figure*}

Figure \ref{fig_cartoon_R25} illustrates how the {\hi} content within the stellar disc is defined for galaxies with different resolutions.
The red circle is the original stellar disc aperture based on $R_{\rm 25}$ and the yellow circle shows the convolved stellar radius used to measure the \hi\ mass enclosed within the stellar disc.
As the angular size of the stellar disc decreases (from left to right), the difference between the red and yellow circles increases due to the beam-smearing effect.

As global {\hi} properties, we derive the global {\hi} mass fraction ($f_{\rm HI}$) and the average {\hi} surface density within the {\hi} disc ($\mu_{\rm HI,RHI}$) as
\begin{equation}
    f_{\rm HI}=\frac{M_{\rm HI}}{M_{\rm \star}},
\end{equation}
\begin{equation}
    \mu_{\rm HI,RHI}=\frac{M_{\rm HI,RHI}}{\pi R_{\rm HI}^2}.
\end{equation}

We derive the {\hi} mass fraction and the average {\hi} surface density within $R_{\rm 25}$ ($R_{\rm 24}$) as
\begin{equation}
    f_{\rm HI,R25(24)}=\frac{M_{\rm HI,R25(24)}}{M_{\rm \star,R25(24)}},
\end{equation}
\begin{equation}
    \mu_{\rm HI,R25(24)}=\frac{M_{\rm HI,R25(24)}}{\pi R_{\rm 25(24),c}^2}.
\end{equation}

\section{Sample selection}
\label{sec_sample}

%\textcolor{red}{ [Here introduce the 2 samples used in this paper: the primary sample with R24>15 arcsec, including xx galaxies, and the secondary sample of xx galaxies with R25>15 arcsec. Use "primary" and "secondary" sample throughout the paper.]}

After measuring {\hi} and stellar properties of 1,543 WALLABY galaxies, we carefully select our sample to ensure accurate measurements of the {\hi} mass within the stellar disc while maximising the number of available WALLABY galaxies.
Measuring \hi\ properties within the stellar disc is meaningful only for galaxies with stellar discs larger than the radio beam, thus we exclude galaxies with stellar discs smaller than one beam.
The parent sample consists of 995 galaxies with $R_{\rm 25}$ larger than half of the beam ($R_{\rm 25}$>15").
The primary sample is a subset of the parent sample, which includes 719 galaxies with $R_{\rm 24}$>15".
We also use \textit{"higher-resolution"} to refer to the subset of galaxies resolved by at least two beams within the stellar disc, defined by $R_{\rm 25}$ (348 galaxies) and $R_{\rm 24}$ (206 galaxies).
In general, results related to $R_{\rm 25}$-based {\hi} within the stellar disc properties are presented using the parent sample, while $R_{\rm 24}$-based properties are presented based on the primary sample, and statistics obtained for the primary sample only are presented where appropriate for comparison.
%Among these, 206 and 348 galaxies (approximately 30\% of each sample) are resolved by at least two beams within the stellar disc defined by $R_{\rm 24}$ and $R_{\rm 25}$, respectively (i.e., better-resolved sample).
Considering the potential uncertainty associated with marginal resolution, as shown in Figure \ref{fig_cartoon_R25}, we present our results using different colours or sizes to indicate the number of ASKAP beams along the major axis of the stellar disc ($R_{\rm 25(24)}/15$").

Figure \ref{fig_sample_RHI} shows the physical properties of 995 galaxies with $R_{\rm 25} > 15$".
We colour-code our galaxies from grey for galaxies with 15"<$R_{\rm 25}$<30" to progressively darker colours for more spatially resolved galaxies.
As expected, the galaxies with better-resolved discs are typically found at low redshifts (top left panel) and large stellar mass (top right panel).
Our {\Mhi} values tend to be higher than the medians for xGASS (\citealp{Catinella2018}; black diamonds), a stellar mass selected \hi\ survey that is not biased toward gas-rich systems, and generally fall below the fitted line from ALFALFA in \citeauthor{Huang2012} (\citeyear{Huang2012}; black dashed line), particularly at low stellar mass, because our sample excludes galaxies with poorer resolution at higher redshifts, which are often low stellar mass and {\hi}-rich.
The bottom panels indicate that distributions of poorly-resolved and well-resolved galaxies overlap well in stellar mass and size or stellar mass and colour plane, and their correlations remain strong.
%\textcolor{red}{Based on the classification of WALLABY \hi\ detections in the Hydra field by \citet{Reynolds2022}, we confirmed that the full and better-resolved subsets ($R_{\rm 25}$>30") contain a comparable fraction of cluster galaxies, suggesting that the potential biases from environmental effects are unlikely to explain the biases from the resolution.}

\section{Results}
\label{sec_result}
\subsection{Global {\hi} vs. {\hi} within the stellar disc}

Before examining how gas scaling relations change when excluding the outer \hi\ regions of galaxies, we first analyse the distribution of \hi\ mass and \hi\ surface density within $R_{\rm 25}$ and $R_{\rm 24}$.
%Previous studies have shown that while outer galaxy regions typically exhibit a uniform profile, {\hi} surface density in the inner regions is more varied \citep{Leroy2008,Wang2014,Eibensteiner2024}.
%We compare global {\hi} properties and {\hi} properties within the stellar disc of WALLABY galaxies.
The top row of Figure \ref{fig_hist_HI_in} presents the distribution of {\hi} mass within the {\hi} disc (i.e., {\hi} with surface density above 1 M$_{\rm \odot}$ pc$^{-2}$) and {\hi} mass within the stellar discs defined by $R_{\rm 25}$ and $R_{\rm 24}$ normalised by the total {\hi} mass.

Panel (a) shows that the fraction of the total \hi\ mass enclosed within $R_{\rm HI}$ varies mostly from 0.8 to 1.0, with medians of 0.89 and 0.92 for the full and higher-resolution samples, respectively.
This indicates that nearly 90\% of the {\hi} has an (inclination uncorrected) average {\hi} surface density above 1 M$_{\rm \odot}$ pc$^{-2}$.
Galaxies with low fractions of \hi\ mass within $R_{\rm HI}$ possibly have disturbed {\hi} discs.
Panels (b) and (c) reveal that the {\hi} content within the stellar disc is more broadly distributed than that within the {\hi} disc.
On average, 68\% of the {\hi} resides within $R_{\rm 25}$ (with individual variation from $\sim$20\% to 100\%) and 54\% within $R_{\rm 24}$ (spanning a range from $\sim$5\% to 100\%).

\begin{figure*}[hbt!]
\centering
\includegraphics[width=0.97\linewidth]{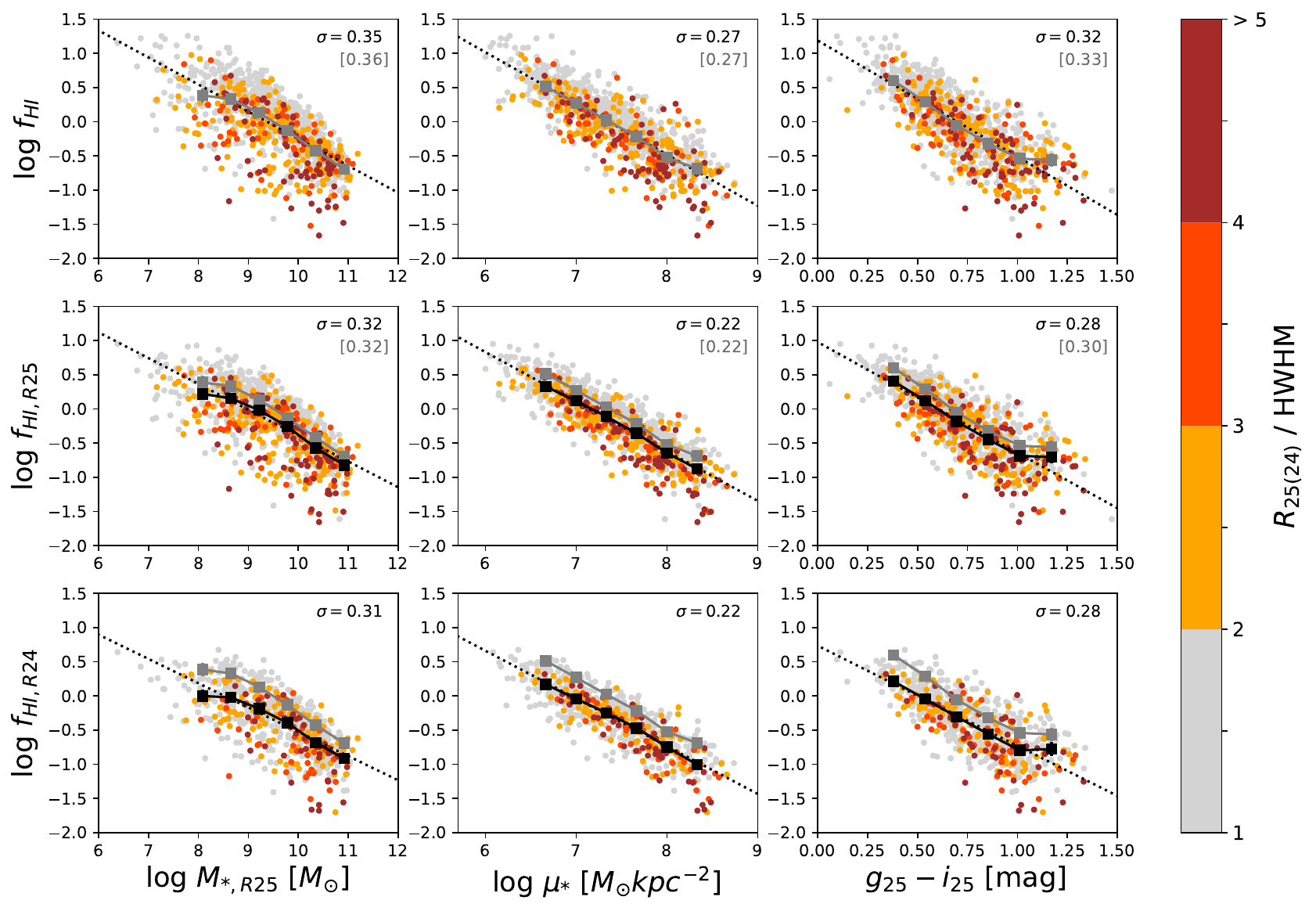}
\caption{Scaling relations of the global {\hi} mass fraction (top row; $f_{\rm HI}=M_{\rm HI}/M_{\rm \star}$) and {\hi} mass fraction within $R_{\rm 25}$ (middle row; $f_{\rm HI,R25}=M_{\rm HI,R25}/M_{\rm \star,R25}$) and $R_{\rm 24}$ (bottom row; $f_{\rm HI,R24}=M_{\rm HI,R24}/M_{\rm \star,R24}$) as a function of stellar mass ($M_{\rm \star,R25}$), stellar mass surface density ($\mu_{\rm \star}$), and {\gi} colour (columns from left to right, respectively) for the parent (top and middle rows) or primary samples (bottom row). The black squares represent the average of logs in each bin, with error bars indicating the standard error of the mean, which is smaller than the symbol size. The grey squares in the top row are replotted in the middle and bottom rows for comparison. The dotted lines show the linear regression fits and the scatter (i.e., the standard deviation along the y-axis from each fitted line) is shown in the top right corner of each panel. The scatter for the primary sample is provided in brackets. Galaxies are colour-coded as in Figure \ref{fig_sample_RHI}, but resolutions are defined based on $R_{\rm 25}$ for $f_{\rm HI}$ and $f_{\rm HI,R25}$ (top and middle rows) and $R_{\rm 24}$ for $f_{\rm HI,R24}$ (bottom row).}
\label{fig_fHI_relations}
\end{figure*}

The bottom row of Figure \ref{fig_hist_HI_in} presents the distributions of the average {\hi} surface density ($\mu_{\rm HI}$) within the \hi\ and stellar discs.
Comparing panels (d), (e), and (f) shows that the average {\hi} surface density within $R_{\rm 25}$ and $R_{\rm 24}$ has a higher median value and a broader distribution than that within $R_{\rm HI}$ (for instance, the Kolmogorov-Smirnov test statistic between the grey distributions in panels d and e is 0.4).
The median values of $\mu_{\rm HI}$ within $R_{\rm HI}$, $R_{\rm 25}$, and $R_{\rm 24}$ are 2.2, 3.0, and 3.7 {\msunpc}, respectively, for the full sample (grey).
The distributions of $\mu_{\rm HI}$ within $R_{\rm 25}$ and $R_{\rm 24}$ extend from 1 to $\sim$6 {\msunpc} for the most part, whereas $\mu_{\rm HI,RHI}$ ranges mostly between 1 and $\sim$4 {\msunpc}.
Few galaxies have average {\hi} surface density within $R_{\rm 25}$ or $R_{\rm 24}$ greater than 9 M$_{\rm \odot}$ pc$^{-2}$, which is the approximate threshold where the conversion from atomic to molecular gas is known to occur \citep[e.g.][]{Martin2001,Bigiel2008}.

In summary, we confirm the prominent variations of {\hi} properties within the stellar disc compared to global {\hi} properties using WALLABY galaxies, which has been suggested in previous studies  \citep[e.g.][]{Leroy2008,Wang2014,Eibensteiner2024} but is now confirmed with a statistical number of galaxies.
We investigate the causes for these broad variations in the following sections.

\begin{figure*}[hpt!]
\centering
\includegraphics[width=0.93\linewidth]{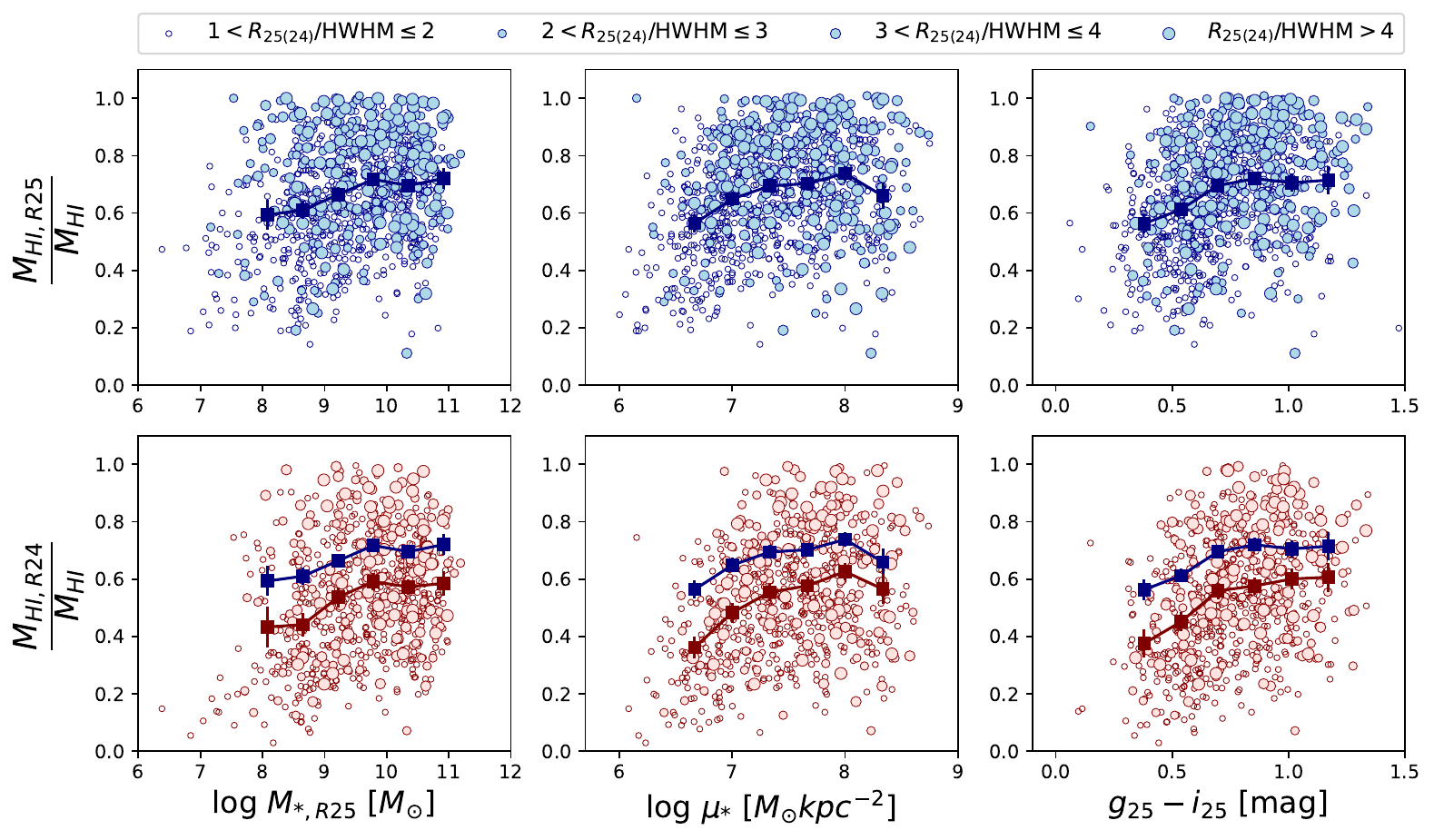}
\caption{The {\hi} mass within $R_{\rm 25}$ (top row) and $R_{\rm 24}$ (bottom row) normalised by the total {\hi} mass is plotted as a function of stellar mass ($M_{\rm \star,R25}$), stellar mass surface density ($\mu_{\rm \star}$), and {\gi} colour (columns from left to right, respectively). Galaxies with stellar disc major axes between 1 and 2 beams (1 < $R_{\rm 25(24)}$/HWHM < 2) are shown as empty circles, while better resolved galaxies are indicated with larger filled circles. Resolutions are defined based on $R_{\rm 25}$ for $M_{\rm HI,R25}/M_{\rm HI}$ (top row) and $R_{\rm 24}$ for $M_{\rm HI,R24}/M_{\rm HI}$ (bottom row). The binned means are represented as squares with error bars that show the standard error of the mean. The means in the top row are replotted as the blue squares in the bottom row for comparison.}
\label{fig_MHI_relations}
\end{figure*}

\subsection{\hi\ scaling relations restricted to $R_{\rm 25}$ and $R_{\rm 24}$}
%\subsection{The fraction of {\hi} mass within the stellar disc}

We examine here the impact of excluding the outer {\hi} regions on gas mass fraction scaling relations. 
Figure \ref{fig_fHI_relations} shows the relation between the {\hi} mass fraction and various stellar properties: stellar mass ($M_{\rm \star,R25}$), stellar surface density ($\mu_{\rm \star}$), and optical colour ({\gi}), comparing the global {\hi} mass fraction ($f_{\rm HI}$; top row) and {\hi} mass fraction confined to $R_{\rm 25}$ ($f_{\rm HI,R25}$; middle row) and $R_{\rm 24}$ ($f_{\rm HI,R24}$; bottom row).
Note that we present the scaling relations as a function of $R_{\rm 25}$-related stellar properties (e.g., $M_{\rm \star,R25}$ and {\gi}) for convenience, but using $R_{\rm 24}$-related properties does not alter the results significantly.
Squares represent the average of the logarithm of the {\hi} mass fraction, ensuring a minimum of 10 galaxies per bin.
Error bars indicate the standard error of the mean but are smaller than the symbol size.
Linear regression fits are included as dotted lines and the scatter (i.e., the standard deviation of the logarithm of the {\hi} mass fraction from the fit) is noted in the top-right corner of each panel.
Table \ref{tab_fHI_relations} provides details of the linear fit parameters, scatter, and the Pearson coefficient ($\varrho$).
As in Figure \ref{fig_sample_RHI}, galaxies are colour-coded by the number of beams along the stellar major axis, i.e., $R_{\rm 25}/15$" ($R_{\rm 24}/15$" for $f_{\rm HI,R24}$ scaling relations), from grey to darker shades.

The global $f_{\rm HI}$ scaling relations (top row) show the well-known trends of decreasing gas mass fraction for galaxies with higher stellar mass, higher stellar surface density, and redder colours.
%There is a slight flattening trend in the lower $M_{\rm \star,R25}$ regime \textcolor{red}{($\lesssim10^{8.5}M_{\rm \odot}$),} but the sample is too limited to confirm this trend.
The trends do not change depending on the samples.
The marginally resolved galaxies (grey dots) tend to have higher $f_{\rm HI}$ compared to better resolved ones (coloured dots) because they tend to be {\hi}-rich galaxies detected at high redshift (see Figure \ref{fig_sample_RHI}).

\begin{table}[t!]
\caption{Parameters of the linear least-squares regression fits ($y=ax+b$), the scatter ($\sigma$), and Pearson correlation coefficients ($\varrho$) for the scaling relations of the global {\hi} mass fraction ($f_{\rm HI}$) and {\hi} mass fraction within $R_{\rm 25}$ ($f_{\rm HI,R25}$) for the parent sample and $R_{\rm 24}$ ($f_{\rm HI,R24}$) for the primary sample (see Figure \ref{fig_fHI_relations}). The $\sigma$ and $\varrho$ for $f_{\rm HI,R25}$ for the primary sample are provided in brackets. The p-value of each Pearson correlation is close to zero.}
\centering
\resizebox{\textwidth}{!}{\begin{tabular}{ c c c c c c }
\toprule
\headrow $y$ & $x$ & $a$ & $b$ & $\sigma$ & $\varrho$ \\
\midrule
 log $f_{\rm HI}$ & log $M_{\rm \star,R25}$ & $-0.40\pm0.01$ & $3.7\pm0.12$ & $0.35~[0.36]$ & $-0.70~[-0.68]$ \\
                  & log $\mu_{\rm \star}$ & $-0.75\pm0.02$ & $5.5\pm0.12$ & $0.27~[0.27]$ & $-0.84~[-0.83]$ \\  
                  & {\gi} & $-1.7\pm0.04$ & $1.2\pm0.03$ & $0.32~[0.33]$ & $-0.77~[-0.74]$ \\  
 log $f_{\rm HI,R25}$ & log $M_{\rm \star,R25}$ & $-0.38\pm0.01$ & $3.4\pm0.11$ & $0.32~[0.32]$ & $-0.72~[-0.71]$ \\ 
                            & log $\mu_{\rm \star}$ & $-0.73\pm0.01$ & $5.2\pm0.10$ & $0.22~[0.22]$ & $-0.87~[-0.88]$ \\    
                            & {\gi} & $-1.6\pm0.04$ & $0.98\pm0.03$ & $0.28~[0.30]$ & $-0.79~[-0.76]$ \\  
 log $f_{\rm HI,R24}$ & log $M_{\rm \star,R25}$ & $-0.36\pm0.01$ & $3.0\pm0.13$ & $0.31$ & $-0.70$ \\ 
                            & log $\mu_{\rm \star}$ & $-0.70\pm0.02$ & $4.9\pm0.12$ & $0.22$ & $-0.86$ \\    
                            & {\gi} & $-1.5\pm0.05$ & $0.74\pm0.04$ & $0.28$ & $-0.75$ \\  
\bottomrule
\end{tabular}}
\label{tab_fHI_relations}
\end{table}

The $f_{\rm HI,R25}$ and $f_{\rm HI,R24}$ scaling relations (middle and bottom rows, respectively) demonstrate trends similar to those of $f_{\rm HI}$, but with systematically lower {\hi} mass fractions.
The $f_{\rm HI,R24}$ scaling relations are highly similar to those of $f_{\rm HI,R25}$, but $f_{\rm HI,R24}$ has slightly lower \hi\ mass fractions because of their smaller \hi\ mass enclosed within $R_{\rm 24}$ compared to $R_{\rm 25}$.
The differences between the average $f_{\rm HI}$ (grey squares) and $f_{\rm HI,R25}$ (black squares in the middle panel) are consistent across the range of stellar properties (0.1-0.2 dex).
Interestingly, compared to the global quantities, the $f_{\rm HI,R25}$ scaling relations show tighter correlations with decreased scatter (which is higher than its standard error) and higher (absolute) values of the Pearson coefficient for all stellar properties, but there is no further improvement when moving to $f_{\rm HI,R24}$.
In particular, $f_{\rm HI,R25}$ shows the strongest correlation and the largest decrease of scatter when plotted as a function of the stellar surface density ($\varrho = -0.87$ and $\Delta \sigma = 0.05$).
The higher-resolution subset shows similar trends, i.e., decreased scatter for $f_{\rm HI,R25}$ by 0.04, 0.06, and 0.03 for the stellar mass, stellar surface density, and colour, respectively, indicating that the trend is not significantly affected by either resolution or sample selection.
The reduced scatter for all stellar properties suggests that the {\hi} mass fraction within the stellar disc is more strongly related to stellar properties than the global {\hi} mass fraction, which implies that examining \hi\ within stellar discs offers a clearer understanding of the relationship between \hi\ content and stellar properties.

\begin{figure}[t!]
\centering
\includegraphics[width=\linewidth]{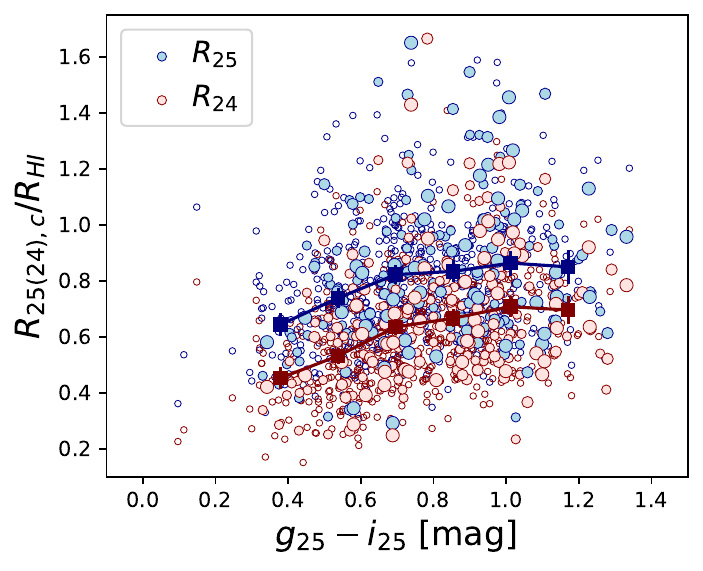}
\caption{The size of the stellar disc relative to the {\hi} disc ($R_{\rm 25,c}/R_{\rm HI}$ in blue circles and $R_{\rm 24,c}/R_{\rm HI}$ in red circles) as a function of {\gi} colour. Markers are the same as in Figure \ref{fig_MHI_relations}. Resolutions are defined based on $R_{\rm 25}$ for $R_{\rm 25,c}/R_{\rm HI}$ (blue circles) and $R_{\rm 24}$ for $R_{\rm 24,c}/R_{\rm HI}$ (red circles).}
\label{fig_R25_g_i}
\end{figure}

\begin{figure*}[t!]
\centering
\includegraphics[width=0.93\linewidth]{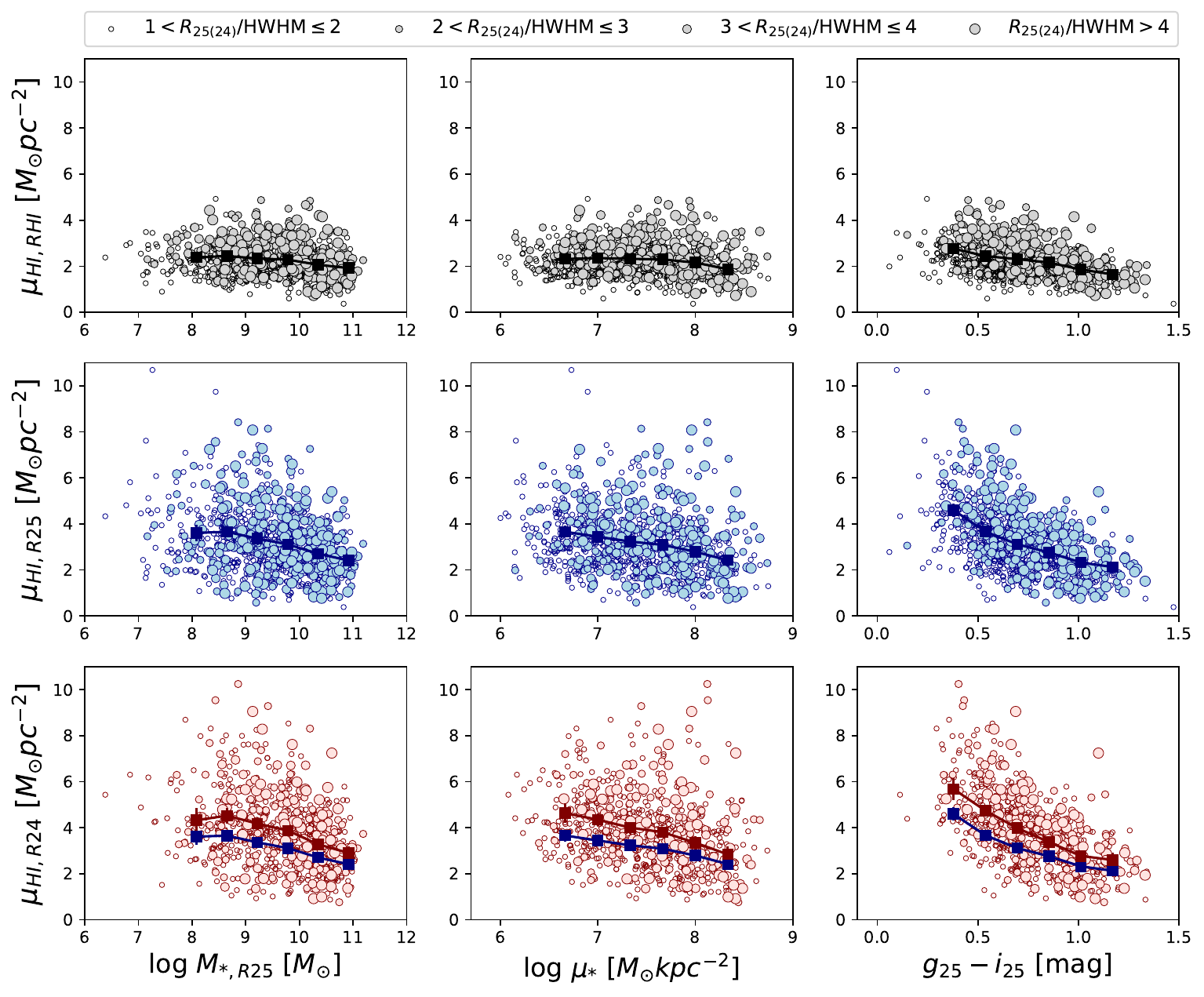}
\caption{The average {\hi} surface density within the {\hi} disc ($\mu_{\rm HI,RHI}$; top row) and within the stellar disc ($\mu_{\rm HI,R25}$ and $\mu_{\rm HI,R24}$; middle and bottom rows, respectively) as a function of stellar mass ($M_{\rm \star,R25}$), stellar mass surface density ($\mu_{\rm \star}$), and {\gi} colour (columns from left to right, respectively). Markers are the same as in Figure \ref{fig_MHI_relations}. Resolutions are defined based on $R_{\rm 25}$ for $\mu_{\rm HI,RHI}$ and $\mu_{\rm HI,R25}$ (top and middle rows) and $R_{\rm 24}$ for $\mu_{\rm HI,R24}$ (bottom row). The means in the middle row are replotted as the blue squares in the bottom row for comparison.}
\label{fig_mu_relations}
\end{figure*}

\subsection{{\hi} mass and surface density within the stellar disc}
\label{sec_result_relations}

In order to gain insights into what drives the decrease in scatter of the \hi\ scaling relations shown in Figure \ref{fig_fHI_relations}, as well as the large variations in the fraction of total \hi\ mass and \hi\ surface density within $R_{\rm 25}$ and $R_{\rm 24}$ observed in Figure \ref{fig_hist_HI_in}, we now explore how these quantities depend on stellar properties.

Figure \ref{fig_MHI_relations} shows the relation between the fraction of the total \hi\ mass enclosed within $R_{\rm 25}$ (top row) or $R_{\rm 24}$ (bottom row) plotted versus stellar mass, stellar surface density, and colour.
Table \ref{tab_MHI_ratio} lists the Spearman coefficients ($\varrho$) for each stellar property.
There are weak positive correlations between the fraction of \hi\ mass within $R_{\rm 25}$ or $R_{\rm 24}$ and these stellar properties ($\varrho\sim 0.3$), which are more pronounced in galaxies with lower stellar mass, lower stellar surface density, and bluer colours.
The correlations are slightly stronger when $M_{\rm HI}$ is restricted to a smaller optical radius, i.e., $R_{\rm 24}$ instead of $R_{\rm 25}$ (e.g., for the primary sample, $\varrho$ for colour increases from 0.19 to 0.33).
The mild positive correlation with colour -- strongest among the given stellar properties -- suggests that bluer galaxies tend to have a smaller fraction of their {\hi} mass within their stellar discs.
Given that {\hi} profiles typically exhibit exponential declines in their outer regions \citep[e.g.][]{Wang2014,Wang2025}, this implies that bluer galaxies are likely to have more extended {\hi} discs relative to their stellar discs, a trend directly shown in Figure \ref{fig_R25_g_i}.
The $R_{\rm 25,c}/R_{\rm HI}$ values may appear higher than those reported in previous studies \citep[e.g. $R_{\rm 25}/R_{\rm HI}\sim$0.5 in ][]{Wang2013,Reynolds2023}, but this discrepancy is due to the use of $R_{\rm 25,c}$ instead of $R_{\rm 25}$, i.e., the median $R_{\rm 25}/R_{\rm HI}$ is 0.54, consistent with the literature.

Similarly to Figure \ref{fig_MHI_relations}, we investigate how the average {\hi} surface density within stellar discs ($\mu_{\rm HI,R25(24)}$) relates to the stellar properties in Figure \ref{fig_mu_relations}.
We present the average {\hi} surface density within the {\hi} disc ($\mu_{\rm HI,RHI}$) scaling relations in the top row and those within stellar discs defined by $R_{\rm 25}$ and $R_{\rm 24}$ in the middle and bottom rows, respectively, to compare the result from global {\hi} and {\hi} within stellar discs.
The Spearman coefficients of each relation are shown in Table \ref{tab_MHI_ratio}.

We find a weak trend in the average {\hi} surface density within the {\hi} disc ($\varrho\sim0.2$), apart from a stronger dependence on colour ($\varrho = -0.42$ for the parent sample).
Noticeable trends emerge for all stellar properties when the {\hi} is restricted to $R_{\rm 25}$ and $R_{\rm 24}$.
Focusing on the average {\hi} surface density within $R_{\rm 25}$ first, its distribution is twice as broad as that within $R_{\rm HI}$ (see also Figure \ref{fig_hist_HI_in}) and it shows stronger correlations with all stellar properties compared to that within $R_{\rm HI}$ ($\Delta\varrho = 0.07, 0.14,$ and $0.11$ with the stellar mass, stellar surface density, and colour, respectively, for the parent sample).
The strongest correlation is with colour ($\varrho = -0.53$ for the parent sample), with a slightly steeper slope in the bluer colour regime.

Interestingly, these trends become more pronounced when $\mu_{\rm HI}$ is further restricted to $R_{\rm 24}$ instead of $R_{\rm 25}$ ($\varrho$ increases by $0.06, 0.05,$ and $0.06$ with the stellar mass, stellar surface density, and colour, respectively, for the primary sample).
This indicates that bluer galaxies, which are likely to be more actively star-forming, tend to have higher {\hi} surface densities within the stellar disc.
Moreover, this trend strengthens as we focus on the inner {\hi} region.
Galaxies with high {\hi} surface density within the stellar disc (> 8 {\msunpc}) tend to be low mass galaxies with high stellar surface density and bluer colour, although their number is small.
In other words, bluer galaxies have simultaneously higher HI surface densities within the stellar disc and more HI reservoir outside the stellar disc than redder systems.

\begin{table}[t]
\caption{Spearman correlation coefficients for $M_{\rm HI,R25}/M_{\rm HI}$, $M_{\rm HI,R24}/M_{\rm HI}$ (see Figure \ref{fig_MHI_relations}), $\mu_{\rm HI,RHI}$, $\mu_{\rm HI,R25}$, and $\mu_{\rm HI,R24}$ (see Figure \ref{fig_mu_relations}) as a function of $M_{\rm \star,R25}$, $\mu_{\rm \star}$, and {\gi} for the parent and primary samples, corresponding to $R_{\rm 25}$- and $R_{\rm 24}$-based properties, respectively. The coefficient for the primary sample is provided in brackets. The p-value of each Spearman correlation is close to zero.}
\centering
\resizebox{\textwidth}{!}{\begin{tabular}{c c c c }
\toprule
\headrow & log $M_{\rm \star,R25}$ & log $\mu_{\rm \star}$ & {\gi}\\
\midrule
$M_{\rm HI,R25}/M_{\rm HI}$ & $0.24~[0.11]$ & $0.27~[0.14]$ & $0.29~[0.19]$\\
$M_{\rm HI,R24}/M_{\rm HI}$ & $0.26$ & $0.33$ & $0.33$\\
\midrule
$\mu_{\rm HI,RHI}$ & $-0.23~[-0.27]$ & $-0.16~[-0.21]$ & $-0.42~[-0.50]$\\
$\mu_{\rm HI,R25}$ & $-0.30~[-0.28]$ & $-0.30~[-0.29]$ & $-0.53~[-0.54]$\\
$\mu_{\rm HI,R24}$ & $-0.34$ & $-0.34$ & $-0.60$\\
\bottomrule
\end{tabular}}
\label{tab_MHI_ratio}
\end{table}

\section{Discussion}
\label{sec_discussion}

In normal star-forming galaxies, it is well known that \hi\ reservoirs extend beyond the inner regions where star formation takes place. This outer \hi\ is less involved in star formation and likely contributes to the large scatter observed in global scaling relations linking gas, stellar, and star formation properties, as well as to the long \hi\ depletion times reported in global studies \citep[e.g.,][]{Catinella2018,Saintonge2022}. With WALLABY we have, for the first time, measured the \hi\ content within the inner regions of galaxies for a statistical sample, allowing us to test some of these expectations. Instead of adopting an arbitrary stellar radius, we compared our results at two optical scales, $R_{\rm 25}$ and $R_{\rm 24}$. These radii approximately correspond to $R_{\rm 90\%}$ for the \iband\ in DECALS imaging and to the radius at 25 mag arcsec$^{-2}$ in the {\Bband}, which has historically been used to define the optical disc \citep[e.g.,][]{Cortese2012}, respectively. For our analysis, we restricted the sample to galaxies with at least one ASKAP beam within $R_{\rm 24}$ (primary sample, 719 galaxies) or $R_{\rm 25}$ (parent sample, 995 galaxies).

Our comparison of \hi\ scaling relations across global and smaller radii reveals the expected, systematic decrease in the \hi\ mass fraction ($M_{\rm HI}/M_{\rm \star}$) at fixed stellar properties, accompanied by a reduction in scatter -- especially at fixed stellar surface density. The decrease in scatter is more pronounced when comparing global measurements to $R_{\rm 25}$ than when comparing $R_{\rm 25}$ to $R_{\rm 24}$. These findings broadly align with some of the results presented by \citet{Wang2020}, who estimated the \hi\ mass fraction within $R_{\rm 90\%}$ (i.e. the radius enclosing 90\% of the total \rband\ flux from SDSS DR7; \citealt{Abazajian2009}) based on spatially unresolved \hi\ data.
They also found a reduced scatter for all given properties, with the largest decrease for the stellar surface density.
Their relations became slightly shallower when the gas mass fraction was restricted to the stellar disc, while we found no significant changes in the slopes.
This discrepancy could stem from variations in selected samples, encompassing different ranges of stellar populations, methodologies -- such as observational-driven models used in their estimates versus our direct measurements -- and definitions of the stellar disc.
%[SUMMARIZE COMPARISON: reduced scatter for all given properties, with the largest decrease for mustar. Note that we find no significant variation in the slopes of these relations, so we need to explain why we differ here.]

Next, we quantified the distributions of \hi\ mass and average \hi\ surface density within $R_{\rm 25}$ and $R_{\rm 24}$, noted the larger variation of these quantities compared to the global ones (Figure \ref{fig_hist_HI_in}), and explored their dependence on stellar properties (Figures \ref{fig_MHI_relations} and \ref{fig_mu_relations}). This showed that the variation in \hi\ drives the most significant change in optical colour. Our main result is Figure \ref{fig_mu_relations}: the global \hi\ surface density is to first order independent of stellar mass and stellar surface density, and shows a slight dependence on colour. This indicates that bluer galaxies tend to have higher average \hi\ surface densities than redder galaxies across their \hi\ discs. When restricted to smaller radii, \hi\ surface densities start showing an anti-correlation with stellar mass and surface density, and the dependence on colour becomes stronger. In other words, the more outer \hi\ (which is not directly involved in star formation) is excluded by going to smaller radii, the stronger the correlation between \hi\ surface density and colour (a proxy for specific star formation) becomes.
These findings are broadly consistent with \citet{Wang2017}, who reported that {\hi}-rich galaxies tend to have higher average \hi\ surface densities within the optical disc, based on a sample of nearby LVHIS galaxies with optical radii larger than two beam sizes (i.e., $\gtrsim$ 2$\times$40 arcsec).

\begin{figure}[t!]
\centering
\includegraphics[width=\linewidth]{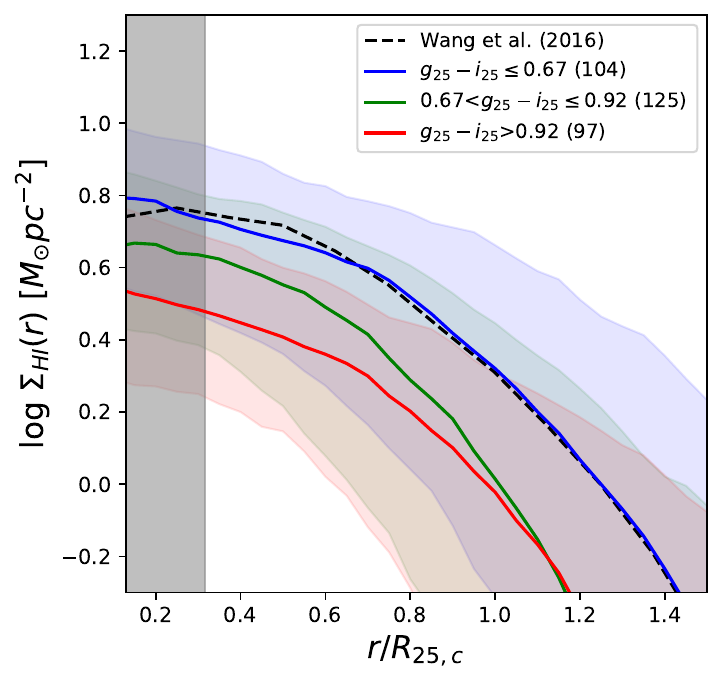}
\caption{The median {\hi} surface density profiles ($\Sigma_{\rm HI}(r)$) as a function of radius scaled by $R_{\rm 25,c}$ binned by colour. The sample only includes the better resolved galaxies with moderate inclination ({\riso} > 30" and $i$ < 80 degrees). The range of each bin and the number of galaxies are shown in the upper right corner. $\Sigma_{\rm HI}(r)$ in each bin are the blue, green, and red solid lines and the shaded regions show the 25th and 75th percentiles. The inclination is corrected using the axes ratio based on the {\iband} image. The median radial profile of 168 late-type galaxies as a function of $R_{\rm HI}$ from \citet{Wang2016} is presented as the black dashed line as a reference, which is scaled by the median ratio of $R_{\rm 25,c}$ to $R_{\rm HI,dep}$ for galaxies with {\gi}$\leq$0.67. The black shaded region is where $R_{\rm 25,c}$ is smaller than 15".}
\label{fig_HI_profile}
\end{figure}

Here we further explore these findings by calculating radial {\hi} surface density profiles binned by colour.
Figure \ref{fig_HI_profile} presents the inclination-corrected median {\hi} surface density profiles ($\Sigma_{\rm HI}(r)$) for a subset of 326 galaxies with {\riso} > 30" and inclination $i$ < 80 degrees, scaled by $R_{\rm 25,c}$.
We note that, since deriving reliable radial {\hi} profiles for WALLABY galaxies is challenging \citep[e.g., marginal resolution, high inclination; see][]{Deg2022}, this figure is not meant to provide a robust quantification of these profiles, but to visually confirm our interpretation.
As can be seen, bluer galaxies have systematically higher \hi\ surface densities both within the stellar disc and on global scales (as shown in Figure \ref{fig_mu_relations}) and have more extended \hi\ discs relative to their stellar discs (as shown in Figures \ref{fig_MHI_relations} and \ref{fig_R25_g_i}).

This is not the first time that trends in radial {\hi} surface density profiles and the relative sizes of \hi\ and stellar discs depending on colour are reported.
\citet{Wang2014} presented median {\hi} radial profiles by categorising 23 galaxies from the Bluedisk survey \citep{Wang2013} based on various {\hi} and stellar properties, identifying $NUV-r$ colour as the property showing the most significant variation.
Our work confirms and quantifies these trends with a much larger sample, using a statistically significant number of galaxies with homogeneously measured {\hi} and stellar properties.
Interestingly, the median {\hi} profile from \citeauthor{Wang2016} (\citeyear{Wang2016}; black dashed line in Figure \ref{fig_HI_profile}), derived from spatially well-resolved observations of 168 spiral and dwarf galaxies, aligns closely with the median {\hi} profile of our galaxies in the blue bin.

The strongest trends with colour suggest a closer connection between the inner gas reservoir and star formation activity in galaxies.
The conversion of gas into stars plays an important role in the evolution of {\hi} discs. 
When the star formation rate surpasses the rate of gas inflow, the stellar disc grows while the gas disc diminishes relative to it.
For instance, \citet{Wang2013} found negative correlations between the relative size of {\hi} disc and both the stellar mass and stellar surface density in Bluedisk galaxies.
\citet{Reynolds2023} reported similar results, but found a stronger correlation with specific star formation rate (sSFR) ($\varrho = 0.45$) than with the stellar mass and stellar surface density ($\varrho = -0.26$ and $-0.23$, respectively) for WALLABY galaxies.
Since bluer galaxies tend to have higher sSFR \citep[e.g.][]{Bigiel2010b}, their findings generally align with our observation of increasing $R_{\rm 25,c}/R_{\rm HI,dep}$ from blue to red galaxies.
This suggests that further gas replenishment may not be significant enough to alter these trends, which needs to be confirmed with the direct measurement of sSFR in a further study.

Interestingly, previous studies of spatially resolved \hi\ scaling relations, based on modest galaxy samples, have shown that the strong correlation between \hi\ and SFR observed on global scales weakens or disappears at sub-kiloparsec scales \citep[e.g.,][]{Wong2002,Boissier2003,Bigiel2008,Schruba2011}.
For example, \citet{Schruba2011}, using 33 galaxies mostly from The \hi\ Nearby Galaxy Survey \citep[THINGS;][]{Walter2008}, found no correlation between \hi\ and SFR surface densities, even in \hi\ dominated regions.
Similarly, \citet{Watts2023} reported weak correlations between \hi\ and stellar surface densities \citep[which correlate with SFR surface densities, e.g.,][]{Morselli2020} for THINGS galaxies, with even weaker correlations for Virgo cluster galaxies \citep[18 galaxies observed in the VLA Imaging of Virgo galaxies in Atomic gas survey, VIVA;][]{Chung2009}, highlighting significant variations between individual systems.
On the other hand, \citet{Bacchini2019} suggested that using deprojected volume densities shows tighter correlations between \hi\ and SFR.
These findings underscore the need for spatially resolved \hi\ studies for statistically significant galaxy samples, to better connect \hi\ gas reservoirs to star formation.

This study is based on a sample that is biased toward gas-rich galaxies and limited in spatial resolution.
To account for the latter, we used different markers to indicate the number of beams along the stellar disc's major axis and demonstrated that the observed correlations remain consistent across both the entire sample and the higher-resolution subset.
Our sample also includes galaxies in dense environments, such as the Hydra cluster where environmental effects may influence their {\hi} content \citep[e.g.,][]{Boselli2006,Cortese2021}.
However, we find that the correlations in the fraction of total \hi\ mass and average \hi\ surface density within the $R_{\rm 25}$ disc scaling relations remain valid even after excluding galaxies with truncated {\hi} discs (i.e., 34 galaxies with $R_{\rm HI}<R_{\rm 25}$).
In conclusion, despite the limitations in spatial resolution and the presence of galaxies in dense environments, our key findings remain robust and highlight the importance of measuring the \hi\ content within the stellar disc to better understand its link with star formation.

\section{Conclusions}
\label{sec_conclusion}

In this paper, we use WALLABY pilot data to quantify, for the first time, \hi\ mass fraction and average \hi\ surface density within the stellar disc, and their dependence on stellar properties, for a statistical sample of galaxies.
We explore how gas scaling relations change as we move from global \hi\ measurements to those restricted to $R_{\rm 25}$, and further to $R_{\rm 24}$. 
Our sample comprises 995 and 719 galaxies, with stellar discs defined by $R_{\rm 25}$ and $R_{\rm 24}$, respectively, resolved by at least one WALLABY synthesised beam.
Our findings are similar even if we limit our sample to galaxies with higher spatial resolution.

\begin{itemize}
    \item On average, about 68\% of the {\hi} is found within the stellar disc defined by $R_{\rm 25}$ (54\% for $R_{\rm 24}$) for WALLABY galaxies. However, the fraction spans a wide range from $\sim$20\% to 100\%, with even greater variation observed for $R_{\rm 24}$. Similarly, the average {\hi} surface density within the stellar disc varies from 1 to 6 {\msunpc}.
    \item The scatter of {\hi} mass fraction ($M_{\rm HI}/M_{\rm \star}$) within the stellar disc scaling relations decreases compared to global {\hi} for all stellar properties, consistent with the findings from spatially unresolved data in \citet{Wang2020}. This suggests a closer connection between \hi\ within the stellar disc and the stars than with global {\hi}.
    \item Notable trends in {\hi} scaling relations become apparent when the {\hi} measurements are restricted to the stellar discs. The strongest correlation is found with colour: bluer galaxies tend to have a lower fraction of their total \hi\ mass, and a significantly higher average {\hi} surface density, within their stellar discs than redder galaxies. These trends become more pronounced when the {\hi} is restricted to smaller radii, implying a stronger link between the inner {\hi} reservoirs and star formation.
    \item The variations in {\hi} properties within stellar discs can be attributed to systematic differences in the radial {\hi} surface density profiles. Bluer galaxies tend to have elevated {\hi} surface densities and larger {\hi} discs relative to their stellar discs than redder galaxies.
\end{itemize}

The WALLABY survey, despite its bias toward gas-rich galaxies and limited spatial resolution, enables us to conduct a statistical analysis that connects {\hi} and stars on the same spatial scale. 
Further research that includes star formation will be important for understanding how {\hi} consumption within the stellar disc, or {\hi} depletion time, differs from the global one and how it relates to galaxy properties.
The full WALLABY survey will provide even more statistically significant insights with an increased number of spatially resolved galaxies.

%\endnote in some journals will behave like \footnote; and \printendnotes will not output anything. 
\printendnotes

\acknowledgement

This scientific work uses data obtained from Inyarrimanha Ilgari Bundara, the CSIRO Murchison Radio-astronomy Observatory. We acknowledge the Wajarri Yamaji People as the Traditional Owners and native title holders of the Observatory site. CSIRO’s ASKAP radio telescope is part of the Australia Telescope National Facility (https://ror.org/05qajvd42). Operation of ASKAP is funded by the Australian Government with support from the National Collaborative Research Infrastructure Strategy. ASKAP uses the resources of the Pawsey Supercomputing Research Centre. Establishment of ASKAP, Inyarrimanha Ilgari Bundara, the CSIRO Murchison Radio-astronomy Observatory and the Pawsey Supercomputing Research Centre are initiatives of the Australian Government, with support from the Government of Western Australia and the Science and Industry Endowment Fund.

WALLABY acknowledges technical support from the Australian SKA Regional Centre (AusSRC).

Parts of this research were supported by the Australian Research Council Centre of Excellence for All Sky Astrophysics in 3 Dimensions (ASTRO 3D), through project number CE170100013.

LC acknowledges support from the Australian Research Council's Discovery Project funding scheme (DP210100337).

JW thanks support of research grants from Ministry of Science and Technology of the People's Republic of China (NO. 2022YFA1602902), National Science Foundation of China (NO. 12073002, 12233001, 8200906879), and the China Manned Space Project.

KS acknowledges support from the Natural Sciences and Engineering Research Council of Canada.

The Legacy Surveys consist of three individual and complementary projects: the Dark Energy Camera Legacy Survey (DECaLS; Proposal ID \#2014B-0404; PIs: David Schlegel and Arjun Dey), the Beijing-Arizona Sky Survey (BASS; NOAO Prop. ID \#2015A-0801; PIs: Zhou Xu and Xiaohui Fan), and the Mayall z-band Legacy Survey (MzLS; Prop. ID \#2016A-0453; PI: Arjun Dey). DECaLS, BASS and MzLS together include data obtained, respectively, at the Blanco telescope, Cerro Tololo Inter-American Observatory, NSF’s NOIRLab; the Bok telescope, Steward Observatory, University of Arizona; and the Mayall telescope, Kitt Peak National Observatory, NOIRLab. Pipeline processing and analyses of the data were supported by NOIRLab and the Lawrence Berkeley National Laboratory (LBNL). The Legacy Surveys project is honored to be permitted to conduct astronomical research on Iolkam Du’ag (Kitt Peak), a mountain with particular significance to the Tohono O’odham Nation.

NOIRLab is operated by the Association of Universities for Research in Astronomy (AURA) under a cooperative agreement with the National Science Foundation. LBNL is managed by the Regents of the University of California under contract to the U.S. Department of Energy.

This project used data obtained with the Dark Energy Camera (DECam), which was constructed by the Dark Energy Survey (DES) collaboration. Funding for the DES Projects has been provided by the U.S. Department of Energy, the U.S. National Science Foundation, the Ministry of Science and Education of Spain, the Science and Technology Facilities Council of the United Kingdom, the Higher Education Funding Council for England, the National Center for Supercomputing Applications at the University of Illinois at Urbana-Champaign, the Kavli Institute of Cosmological Physics at the University of Chicago, Center for Cosmology and Astro-Particle Physics at the Ohio State University, the Mitchell Institute for Fundamental Physics and Astronomy at Texas A\&M University, Financiadora de Estudos e Projetos, Fundacao Carlos Chagas Filho de Amparo, Financiadora de Estudos e Projetos, Fundacao Carlos Chagas Filho de Amparo a Pesquisa do Estado do Rio de Janeiro, Conselho Nacional de Desenvolvimento Cientifico e Tecnologico and the Ministerio da Ciencia, Tecnologia e Inovacao, the Deutsche Forschungsgemeinschaft and the Collaborating Institutions in the Dark Energy Survey. The Collaborating Institutions are Argonne National Laboratory, the University of California at Santa Cruz, the University of Cambridge, Centro de Investigaciones Energeticas, Medioambientales y Tecnologicas-Madrid, the University of Chicago, University College London, the DES-Brazil Consortium, the University of Edinburgh, the Eidgen\"ossische Technische Hochschule (ETH) Zurich, Fermi National Accelerator Laboratory, the University of Illinois at Urbana-Champaign, the Institut de Ciencies de l’Espai (IEEC/CSIC), the Institut de Fisica d’Altes Energies, Lawrence Berkeley National Laboratory, the Ludwig-Maximilians-Universit\"at M\"unchen and the associated Excellence Cluster Universe, the University of Michigan, NSF’s NOIRLab, the University of Nottingham, the Ohio State University, the University of Pennsylvania, the University of Portsmouth, SLAC National Accelerator Laboratory, Stanford University, the University of Sussex, and Texas A\&M University.

BASS is a key project of the Telescope Access Program (TAP), which has been funded by the National Astronomical Observatories of China, the Chinese Academy of Sciences (the Strategic Priority Research Program “The Emergence of Cosmological Structures” Grant \# XDB09000000), and the Special Fund for Astronomy from the Ministry of Finance. The BASS is also supported by the External Cooperation Program of Chinese Academy of Sciences (Grant \# 114A11KYSB20160057), and Chinese National Natural Science Foundation (Grant \# 12120101003, \# 11433005).

The Legacy Survey team makes use of data products from the Near-Earth Object Wide-field Infrared Survey Explorer (NEOWISE), which is a project of the Jet Propulsion Laboratory/California Institute of Technology. NEOWISE is funded by the National Aeronautics and Space Administration.

The Legacy Surveys imaging of the DESI footprint is supported by the Director, Office of Science, Office of High Energy Physics of the U.S. Department of Energy under Contract No. DE-AC02-05CH1123, by the National Energy Research Scientific Computing Center, a DOE Office of Science User Facility under the same contract; and by the U.S. National Science Foundation, Division of Astronomical Sciences under Contract No. AST-0950945 to NOAO.

\bibliography{reference}
%\printbibliography

\clearpage

\end{document}